\documentclass{svproc}
\usepackage{url}
\usepackage{amssymb}
\usepackage{amsmath}
\usepackage{epsfig}

\begin{document}
\mainmatter              
\title{Majorana Dark Matter and Neutrino mass in a singlet-doublet extension of the Standard Model}
\titlerunning{Singlet-Doublet Majorana Dark Matter}  
%
\author{Manoranjan Dutta\inst{1}, Subhaditya Bhattacharya\inst{2},
Purusottam Ghosh\inst{3}, Narendra sahu\inst{1} }
\authorrunning{M Dutta et al.} 
%
%
\institute{Department of Physics, Indian Institute of Technology Hyderabad,\\ Kandi, Telangana-502285, India.\\
\and
Department of Physics, Indian Institute of Technology Guwahati,\\ North Guwahati, Assam-781039, India.\\
\and
Regional Centre for Accelerator-based Particle Physics, Harish-Chandra Research Institute, HBNI,\\ Chhatnag Road, Jhunsi, Allahabad - 211 019, India.\\
\email{ph18resch11007@iith.ac.in, subhab@iitg.ac.in, purusottamghosh@hri.res.in, nshau@phy.iith.ac.in}}

\maketitle              

\begin{abstract}
A minimal extension of the Standard Model (SM) by a vector-like fermion doublet and three 
right handed (RH) singlet neutrinos is proposed in order to explain dark matter and tiny neutrino mass simultaneously. The DM arises as a mixture of the neutral component of the fermion doublet 
and one of the RH neutrinos, both assumed to be odd under an imposed $\mathcal{Z}_2$ symmetry. Being Majorana in nature, the DM escapes from $Z$-mediated direct search constraints to mark a 
significant difference from singlet-doublet Dirac DM. The other two $\mathcal{Z}_2$ even heavy RH neutrinos give rise masses and mixing of light neutrinos via Type-I Seesaw mechanism. 
Relic density and direct search allowed parameter space for the model is investigated through detailed numerical scan.
\end{abstract}
\section{Introduction}
Despite compelling evidences from galaxy rotation curves, gravitational lensing, cosmic microwave background (CMB) etc., we are yet to pin down what dark matter (DM) actually is. Amongst different
class of possibilities, Weekly Interacting Massive Particles (WIMPs), where DM acts as a thermal relic from the early universe and can be probed via both direct and collider search 
experiments is a promising framework. Tiny yet non-zero masses of the neutrinos is another potential mystery, which we simultaneously address in this framework, where the SM has been extended minimally 
with a vector-like fermion doublet and three right handed singlet neutrinos (see~\cite{singlet-doublet_majorana}). 
Right handed neutrino having same $Z_2$ charge as of the vector like doublet, mixes with the neutral component of the 
doublet to render a stable Majorana DM after electroweak symmetry breaking (EWSB). 
The Majorana nature of the DM marks a crucial difference from the singlet-doublet Dirac DM studied earlier~\cite{singlet-doublet_dirac} in relic density and direct search allowed parameter space. It is worthy to note that only singlet or only doublet fermion DM is phenomenologically constrained significantly. 

\section{The Model for Singet-Doublet Majorana Dark Matter}
In this work, the SM is extended by one vector-like fermion doublet (VLFd) $\Psi^T=(\psi^0, \psi^-)$ (with hypercharge 
$Y=-1$ where $Q=T_3+Y/2$) and three heavy right handed singlet neutrinos (RHN) ${N}_{R_i} (i=1,2,3)$. An additional $\mathcal{Z}_2$ symmetry is imposed under which $\Psi$ and $N_{R_1}$ are odd, while all other fields are even. The Lagrangian of the model is given by:
\begin{equation}
\begin{aligned}
\label{model_Lagrangian}
 \mathcal{L} &= \mathcal{L}_{SM} + \overline{\Psi} \left( i\gamma^\mu D_\mu - M \right) \Psi +\overline{{N}_{R_i}} i\gamma^\mu\partial_\mu {N}_{R_i}- (\frac{1}{2}M_{R_i} \overline{{N}_{R_i}} \left({N}_{R_i}\right)^c + h.c) \\
 & -\left[ \frac{Y_1}{\sqrt{2}}\overline{\Psi}\Tilde{H}\big(N_{R_1}+(N_{R_1})^c\big) +h.c\right]- \left(Y_{j \alpha }\overline{N_{R_j}} \Tilde{H^\dagger} L_{\alpha} + h.c.\right).
 \end{aligned}  
\end{equation}
 where $\Tilde{H}=i\tau_2 H^{*}$  and $L$ denotes SM lepton doublet with indices $j=2,3$ and $\alpha = e,\mu, \tau$. ${N}_{R_1}$ being odd under $\mathcal{Z}_2$ has Yukawa coupling ($Y_1$) with fermion doublet 
 $\Psi$ and determines the lightest stable component as DM after EWSB. $N_{R_2}$ and ${N}_{R_3}$ being assumed even under $\mathcal{Z}_2$, don't couple to 
$\Psi$, but couple to the SM lepton doublets ($\sim Y_{j \alpha }$) generating Dirac masses for SM neutrinos after EWSB. The mass terms for the dark sector (after EWSB) can be written as:
\begin{equation}
  -\mathcal{L}_{mass} = M\overline{\psi^0_L}\psi^0_R + \frac{1}{2}M_{R_1}\overline{N}_{R_1}(N_{R_1})^c + \frac{m_D}{\sqrt{2}} (\overline{\psi^0_L}N_{R_1}+\overline{\psi^0_R}(N_{R_1})^c) + h.c. 
  \label{l_mass}
\end{equation}
where $m_D=\frac{{Y_1 v }}{\sqrt{2}}$, where $ v = 246$ GeV. In  
$ ((\psi^0_R)^c, \psi^0_L, (N_{R_1})^c)^T$ basis, the mass matrix becomes:
\begin{equation}\label{dark-sector-mass}
\mathcal{M}=
\left(
\begin{array}{ccc}
0 &M &\frac{m_D}{\sqrt{2}}\\
M &0 &\frac{m_D}{\sqrt{2}}\\
\frac{m_D}{\sqrt{2}} &\frac{m_D}{\sqrt{2}} &M_{R_1}\\
\end{array}
\right)\,.
\end{equation}
The symmetric matrix $\mathcal{M}$ can be diagonalised by $\mathcal{U}.\mathcal{M}.\mathcal{U}^T = \mathcal{M}_{Diag.}$, where, 
\begin{equation}
\label{diagonalizing_matrix}
\mathcal{U}= \left(
\begin{array}{ccc}
1 & 0 & 0\\
0 & e^{i\pi/2} & 0\\
0 & 0 & 1\\
\end{array}
\right)
\left(
\begin{array}{ccc}
\frac{1}{\sqrt{2}}\cos\theta &\frac{1}{\sqrt{2}}\cos\theta &\sin\theta\\
-\frac{1}{\sqrt{2}} &\frac{1}{\sqrt{2}} &0\\
-\frac{1}{\sqrt{2}}\sin\theta &-\frac{1}{\sqrt{2}}\sin\theta &\cos\theta\\
\end{array}
\right)\\.
\end{equation}
The extra phase matrix is multiplied to make sure all the eigenvalues are positive. The diagonalisation of the mass matrix (Eq. \ref{dark-sector-mass}) requires $\tan2\theta = \frac{2 m_D}{M-M_{R_1}}$.
All the three physical states $\chi_{_1}, \chi_{_2} ~{\rm and} ~\chi_{_3}$ are therefore of Majorana nature and their 
mass eigenvalues in the small mixing limit ($\theta\to 0$) are given by,
\begin{equation}
 ~~ m_{\chi_{_1}}  \approx M + \frac{ m^2_D}{M - M_{R_1}}, m_{\chi_{_2}}  = M, ~ m_{\chi_{_3}}  \approx M_{R_1} - \frac{m^2_D}{M - M_{R_1}}~~,
 \label{eq:parameters}
\end{equation}
where we have assumed $m_D << M, M_{R_1}$. It is clear that $m_{\chi_{_1}}>m_{\chi_{_2}}>m_{\chi_{_3}}$ 
and therefore $\chi_{_3}$ becomes the stable DM candidate. The phenomenology of dark sector is therefore governed by the three independent parameters {\it viz.} the DM mass ($m_{\chi_{_3}}$), its mass splitting with the 
heavier neutral component ($\Delta M=m_{\chi_{_1}}-m_{\chi_{_3}} \approx m_{\chi_{_2}}-m_{\chi_{_3}}$), and the doublet-singlet mixing ($\sin\theta$).

\section{Relic Abundance and Direct Search prospects}
\label{relics}

\begin{figure}[htb!]
$$
\includegraphics[height=4.1cm]{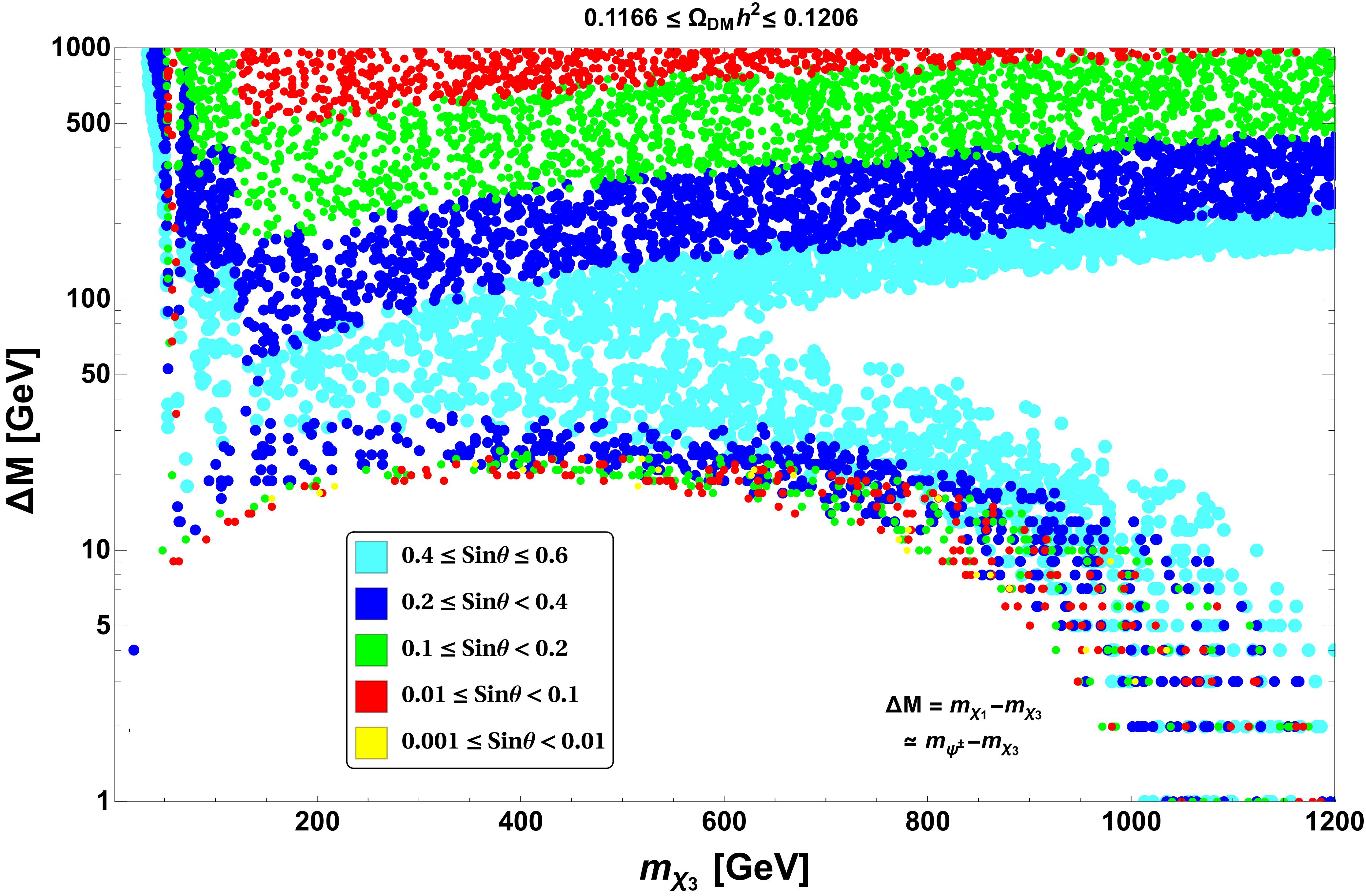}
$$ 
 \caption{\footnotesize{ DM relic density ($0.1166\leq\Omega_{DM} h^2\leq 0.1206$) allowed parameter space in the plane of $\Delta M $ vs $m_{\chi_3}$ for the singlet-doublet Majorana DM model.}}
 \label{relic}
\end{figure}

DM ($\chi_3$) has both gauge interactions (via doublet) and Higgs portal interactions (via $Y_1$) with SM, which keeps it in thermal bath in the early universe and thereafter `{\it freezes out}' via the number changing processes 
to provide the correct DM relic density ($\Omega h^2 \simeq 0.1$), provided by PLANCK data~\cite{planck}. Using {\tt MicrOmegas}~\cite{micromegas}, we calculate relic density allowed parameter space of the model, shown in Fig.~\ref{relic}, in the $\Delta M - m_{\chi_{_3}}$ plane for different $\sin\theta$ ranges (see figure inset). The region with smaller $\Delta M$ has larger co-annihilation contribution like $\chi_3\chi^\pm \to SM$, a crucial feature of singlet-doublet frameworks. 

\begin{figure}[htb!]
$$
\includegraphics[height=4.0cm]{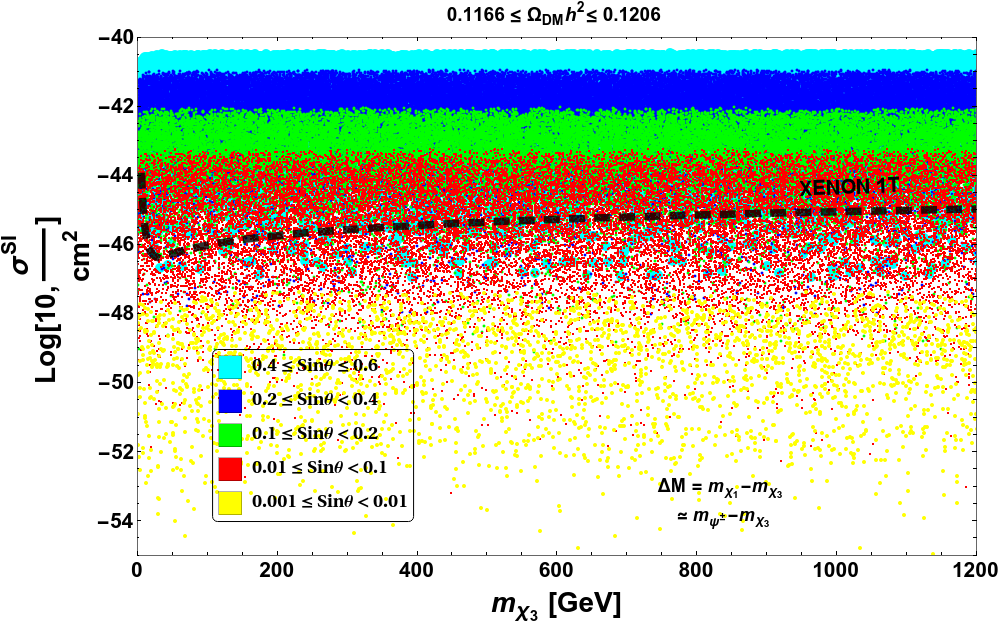}
 \includegraphics[height=4.0cm]{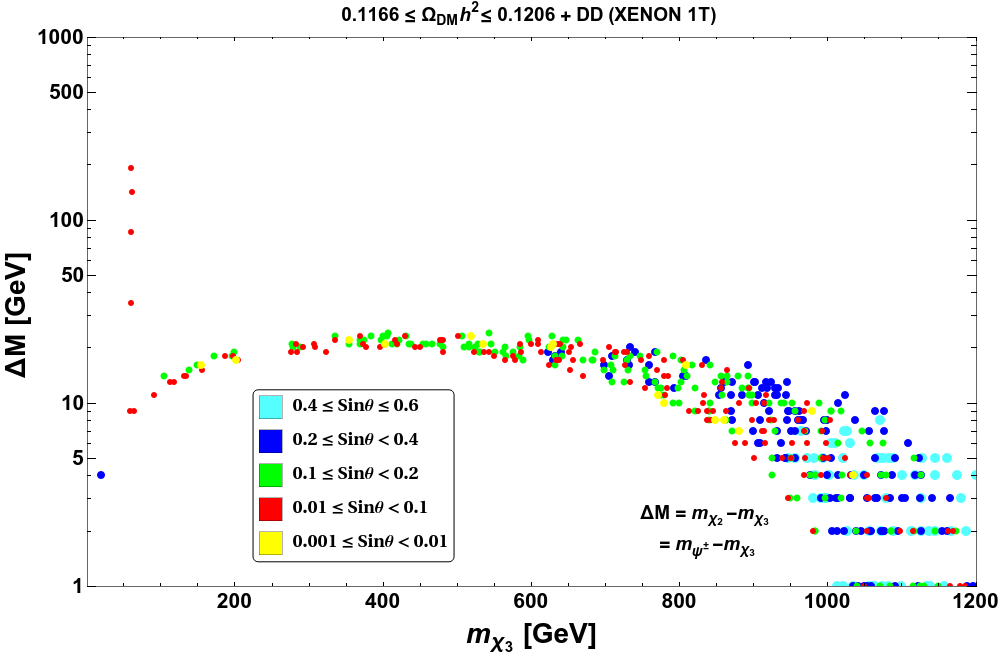}  
$$ 
\caption{\footnotesize{[Left]: Direct detection cross section for the DM ($\chi_3$) confronted with bounds on spin-independent elastic scattering cross section by XENON-1T;
 [Right]: Correct DM relic density in $\Delta M-m_{\chi_{_3}}$ plane constrained by XENON-1T bound. }}
 \label{directdetection}
\end{figure}

Direct detection of DM is possible through elastic scattering of the DM with detector nuclei via Higgs-portal interaction ($Y_1$). Being a Majorana fermion, the DM does not have Z-mediated elastic scattering. 
In the left panel of Fig.~\ref{directdetection}, we confront the direct detection cross section for the model as a function of DM mass with XENON-1T data (shown by black dashed curve)~\cite{xenon1t}. 
The points allowed by both relic and direct search constraints are shown in the right panel of Fig.~\ref{directdetection}, in the $\Delta M-m_{\chi_{_3}}$ plane. We see that 
direct search constraints limits $\Delta M$ (upto 10-15 GeV, excepting resonance region, $m_{\chi_3}\sim m_Z/2$) and mixing $\sin\theta\le 0.6$ as Higgs portal coupling $Y_1 = \frac{\sqrt{2}\Delta M \sin2\theta}{v}$. 
Somewhat larger mixing with doublet ($\sin\theta \sim 0.6$) as allowed here, crucially distinguishes the model from singlet-doublet Dirac DM case, where $\sin\theta\sim 0.05$~\cite{singlet-doublet_dirac} 
is much smaller due to the presence of $Z$ mediated direct search interaction.  

\section{Non-Zero Neutrino mass}
\label{neutrino}
A tiny neutrino mass is generated via Type I seesaw from parts of Eqn.\ref{model_Lagrangian},
\begin{equation}
   - \mathcal{L}^{\nu}_{mass} \supset \Big(Y_{j \alpha }\overline{N_{R_j}} \Tilde{H^\dagger} L_{\alpha} + h.c.\Big) + \Big( \frac{1}{2} M_{R_j}\overline{N}_{R_j}(N_{R_j})^c + h.c.\Big).
\end{equation}where $\alpha = e,\mu, \tau$ and $j = 2,3$.
\newpage
In the basis where the the heavy Majorana mass matrix that takes part in 
seesaw is diagonal {\it i.e.,} $M_{R} = {\rm Diag}(0, M_{R_2}, M_{R_3})$, the light neutrino mass matrix obtained through Type-I seesaw is given by $ m_\nu  = - m_D M^{-1}_R m^T_D$.

\section{Conclusion}
\label{conclusion}
In an arguably simplest extension of SM with a vector like doublet ($\Psi$) and three RHNs ($N_{R_i}$), one achieves a stable Majorana fermion DM out of doublet-singlet mixing having same $\mathcal{Z}_2$ charge, 
and correct neutrino mass via Seesaw I mechanism. The relic density and direct search allowed parameter space of the model allows one to search for the model at Large Hadron Collider with 
leptonic signature accompanied with missing energy or with disappearing charge track and also provides a possible distinction with singlet-doublet Dirac DM through the mixing parameter ($\sin\theta$).      
%
%


\begin{thebibliography}{6}
%
\bibitem{singlet-doublet_majorana} 
M. Dutta, S. Bhattacharya, P. Ghosh and N. Sahu, {\bf JCAP03(2021)008}, [arXiv:2009.00885].

\bibitem{singlet-doublet_dirac}
S. Bhattacharya, N. Sahoo, and N. Sahu, {\bf Phys. Rev. D 96 (2017)}, no. 3 035010, [arXiv:1704.03417];
S. Bhattacharya, P. Ghosh, and N. Sahu, {\bf JHEP 02 (2019) 059}, [arXiv:1809.07474];
S. Bhattacharya, N. Sahoo, and N. Sahu, {\bf Phys. Rev. D 93 (2016)}, no. 11 115040, [arXiv:1510.02760];
S. Bhattacharya, P. Ghosh, N. Sahoo, and N. Sahu, {\bf Front. in Phys. 7 (2019)} 80,
[arXiv:1812.06505].

\bibitem{planck} 
 {\bf Planck} Collaboration, P. Ade et al., {\bf Astron. Astrophys. 571 (2014) }, A16, [arXiv:1303.5076].

\bibitem{micromegas} G. Belanger, F. Boudjema, A. Pukhov, and A. Semenov, {\bf Comput. Phys. Commun. 180 (2009)} 747-767, [arXiv:0803.2360].

\bibitem{xenon1t} 
{\bf XENON} Collaboration, E. Aprile et al., {\bf Phys. Rev. Lett. 121 (2018)} no. 11 111302, [arXiv:1805.12562].





\end{thebibliography}
\end{document}